\documentclass{JHEP3}

\usepackage{amsmath}
\usepackage{amssymb}
\usepackage{graphicx}

\newcommand{\pd}{\partial}

\newcommand{\tg}{\tilde{g}}
\newcommand{\tD}{\tilde{D}}
\newcommand{\tBox}{\tilde{\Box}}
\newcommand{\tR}{\tilde{R}}
\newcommand{\tG}{\tilde{G}}

\newcommand{\tF}{\tilde{F}}

\newcommand{\tX}{\tilde{X}}
\newcommand{\tV}{\tilde{V}}
\newcommand{\tM}{\tilde{M}}
\newcommand{\tLambda}{\tilde{\Lambda}}
\newcommand{\tGamma}{\tilde{\Gamma}}

\newcommand{\diag}{\mathop{\mathrm{diag}}\nolimits}

\newcommand{\const}{\text{const}}

\def\APJ{ Astroph.~J.~}
\def\AJ{ Astron.~J.~}

\title{Towards a covariant model for cosmic self-acceleration}

\author{Alexey S. Koshelev\footnote{On leave from
\textit{Steklov Mathematical Institute of RAS, Gubkin st., 8,
119991, Moscow, Russia, E-mail} \texttt{koshelev@mi.ras.ru}}~~and Theodore N. Tomaras\\
Department of Physics and Institute of Plasma Physics, University of
Crete, GR-710 03 Heraklion, Crete, Greece,\\
E-mails: \email{koshelev@physics.uoc.gr},
\email{tomaras@physics.uoc.gr}}

\abstract{An explicitly covariant formulation is presented of a
modified DGP scenario proposed recently \cite{hep-th/0612213}, to
avoid the instability of the self-accelerating branch. It is based
on the introduction of a bulk scalar field with appropriate
non-minimal coupling to the bulk Einstein-Hilbert term. The method
is general and may be applied to other models as well. }

\preprint{}


\begin{document}


\section{Introduction}

The current observational data \cite{cosm_data} strongly support
Dark Energy domination in the Universe. The nature of the Dark
Energy is still a mystery and the ``coincidence question''
unanswered. So, alternative pictures exist in the literature, among
which is the one proposed recently in the framework of Brane-World
cosmology \cite{AHDD,RS,DGP}, based on the observation that due to
energy exchange with the bulk, the present accelerating Universe may
be a late-time stable attractor of the cosmological evolution
equations \cite{kpt}. This is an effort towards a natural resolution
of the ``coincidence problem'' of cosmology, which in addition leads
to several phenomenologically interesting properties at the fixed
point. However, to accommodate the correct amount of matter
$\Omega_m \simeq 0.3$ at the fixed point, one has to assume that the
brane has negative tension, with the exponential expansion on the
brane driven by energy influx from the bulk \cite{kkttz}.

However, negative tension branes are believed a priori to be
unsatisfactory for a realistic physical model, for essentially two
reasons. First, it is known that the gravitational force on a
negative tension brane in Randall-Sundrum-like scenarios \cite{RS}
(with AdS bulk) becomes repulsive. Indeed, following
\cite{gr-qc/9910076} one can show that the four-dimensional Newton's
constant $G_N$ in this case is proportional to the brane tension
$\sigma$ with positive definite coefficient, leading to negative
$G_N$ for negative $\sigma$. Additional matter in the bulk, for
instance in the form of a minimally coupled scalar field with normal
kinetic term and arbitrary potential,  does not improve the
situation \cite{hep-th/0602149}. Second, negative tension RS-II
branes are believed to be unstable.

However, the physics changes considerably, if one extends the model
to include the Ricci scalar term on the brane, induced by matter
quantum loops or by finite brane-thickness effects. The action of
this, so called DGP, model \cite{DGP} in its simplest version
(without bulk cosmological constant or matter) is
\begin{equation}
S=\frac12\int d^4xdy\sqrt{-\tg}  \tM^3 \tR
+\int d^4x\sqrt{-g}\tM^3\Delta K +{\frac12}\int d^4x\sqrt{-g} M^2 R+\sigma\int d^4x\sqrt{-g}
\label{pure_DGP}
\end{equation}
and describes a 3-brane with tension $\sigma$ but no extra matter on
it, embedded in Minkowski 4+1-dimensional bulk. $R$ is the
aforementioned intrinsic scalar curvature term on the brane, while
$\Delta K=K^+-K^-$ is the jump of the trace of the extrinsic
curvature across the brane. As usual, the purpose of this modified
$K$ dependent Gibbons-Hawking (GH) term is to cancel unwanted terms
in the variation of the bulk action related to the discontinuities
of the derivative of the metric across the brane, so that one ends
up with the proper Israel matching conditions\footnote{A very
explicit derivation of the variation of the GH term can be found in
\cite{chamblin_reall}.}.

The solution for the metric, relevant to cosmology which is our main
interest here, is found to be
\begin{equation}
\tg_{AB}=\left(\begin{tabular}{cc}$(1+\varepsilon
H|y|)^2\left(\begin{tabular}{cc}$-1$&$0$\\$0$&$e^{2Ht}\delta_{ab}$\end{tabular}\right)$&$0$\\$0$&1
\end{tabular}\right)
\label{sol_DGP}
\end{equation}
where $Z_2$ symmetry across the brane is manifest and
$\varepsilon=\pm1$. The parameter $H$ is given by
\begin{equation}
H^2=2\varepsilon\frac{\tM^3}{M^2}H+\frac{\sigma}{3M^2}~\Rightarrow~
H=\frac{\tM^3}{M^2}\left(\varepsilon\pm\sqrt{1+\frac{\sigma
M^2}{3\tM^6}}\right). \label{Hwithsigma}
\end{equation}
Metrics with opposite $H$ are related by time reversal. So, only two
of these four values of $H$ correspond to independent solutions. We
choose the ones which correspond to the two values of
$\varepsilon=\pm 1$, both with the $+$ sign in front of the square
root.

We are interested in solutions with the following properties: (a)
$H>0$, in order to correctly describe the present accelerating
expansion of the Universe, (b) $\varepsilon H<0$ in order to improve
the chances for stability, (c) attractive Newton's law on the brane,
and (d) all the above consistent even with $\sigma\leq 0$. This, may
eventually allow for a natural explanation of the cosmic
acceleration and a resolution of the ``coincidence issue''
\cite{kpt}.

Notice, though, that none of the solutions given in (\ref{sol_DGP}),
(\ref{Hwithsigma}) satisfies these requirements. Closest to being
satisfactory is the self-accelerating one, corresponding to
$\varepsilon=+1$. This has $H>0$ even with a not too negative
$\sigma$, but the metric in this case grows in the bulk and leads,
perhaps not surprisingly, to ghost instabilities in the spectrum of
perturbations \cite{effective1, effective2, Koyama1, Koyama2,
Carena, hep-th/0604086}. For a positive tension brane a helicity-0
excitation of the spin-2 graviton is a ghost, while for a negative
tension brane the spin-0 mode becomes a ghost. For tensionless brane
the ghost field is a linear combination of the spin-0 mode and the
helicity-0 excitation of the graviton.

The other independent solution with $\varepsilon=-1$ is stable, but
it is unsatisfactory in connection with (d) above, because one needs
positive cosmological constant on the brane to explain the
accelerating cosmic expansion. Is there a way to satisfy all four
requirements (a)-(d)?

A step in this direction was outlined in
\cite{hep-th/0612213}\footnote{For a related approach see also
\cite{sahni}.}. The proposal was to modify the bulk action
(\ref{pure_DGP}) by the multiplication with an appropriate smearing
function near the brane as follows
\begin{equation}
S_{\text{bulk}}={\frac12}\int d^4xdy\sqrt{-\tg} \tM^3 \tR \to
{\frac12}\int d^4xdy\sqrt{-\tg} \tM^3 \tF(x,y) \tR
\label{Sgabadadge}
\end{equation}
where in the simplest case $\tF(x,y)=1-m\bar\delta(y)$ with $m$ a
parameter and $\bar\delta(y)$ a $\delta$-like function with a second
parameter $\alpha$. A simple choice is
$\bar\delta(y)=\pi^{-1}\alpha/(\alpha^2+y^2)$. The extrinsic
curvature terms should be modified accordingly. The model still has
the solution (\ref{sol_DGP}) for the metric, with $H$ now given by
\begin{equation}
H=\frac{\tM^3}{M^2}\left(1-\frac{m}{\pi\alpha}\right)\left(\varepsilon\pm\sqrt{1+\frac{\sigma
M^2}{3\tM^6\left(1-\frac{m}{\pi\alpha}\right)^2}}\right).
\label{Hgabadadge}
\end{equation}

The solution with the $+$ sign in front of the square root,
$\varepsilon=-1$ and in the limit $\alpha\to 0$ and $m\to 0$ with
$m/(\pi\alpha)\sim\const>1$, has $H>0$, describes a self-accelerated
brane, which satisfies all our requirements, with the exception of
the stability issue which cannot be decided at the level of such a
non-covariant formulation of the model. Nevertheless, as it was shown in
\cite{hep-th/0612213} such a modification leads to a flip
of the sign in front of the extrinsic curvature terms in the brane equations of motion. Exactly this sign is responsible for the stability
of the metric perturbations and the absence of unstable modes in the normal branch.
Thus, modifying the action in the above mentioned way
we expect to obtain self-acceleration while keeping equations as they are in the normal stable branch. Further nice properties of
this modification are considered the relaxation of the bulk gravity scale and the conservation of the number of parameters.

The purpose of the present paper is to develop a manifestly
covariant formulation of the above modification, making use of a bulk scalar field instead of the function $\tF(x,y)$. Naturally, extra complications arise from the fact that one has to satisfy also the scalar field equation of motion, as well as the corresponding additional matching condition on the brane.


\section{Codimension-1 brane in 5-dimensional bulk}

Consider a 3-brane embedded in a 5-dimensional bulk. The coordinates
in the bulk are denoted by $x^{A}$ with capital latin indices
running from 0 to 4 and with $x^4\equiv y$. The coordinates on the
brane are denoted by $\xi^{\mu}$, with Greek indices from the middle
of the alphabet taking values from 0 to 3. Occasionally, we shall
use $t$ instead of the coordinate with index $0$, while spatial
indices on the brane will be denoted by lowercase Latin letters $a,
b, \dots$. The position of the brane in the bulk is parameterized as
$x^{A}=X^{A}(\xi^{\mu})$, $X^4\equiv Y$. Finally, we shall be using
tildes to designate quantities referring to the bulk. Thus, the bulk
metric is $\tg_{AB}$ and the induced metric on the brane is
$g_{\mu\nu}=\pd_{\mu}X^{A}\pd_{\nu}X^{B}\tg_{AB}(X^{A}(\xi^\mu))$.
Here $\pd_\mu \equiv \pd /\pd\xi^\mu$ and our convention for the signature is
$(-,+,+,+,+)$.

We shall be interested in the model described by the generic action
\begin{eqnarray}
S_{\text{bulk}}&=&\int
d^4xdy\sqrt{-\tg}\left(\frac{\tM^3}2\tF(\Phi)\tR-\frac12\tX(\Phi)\tg^{AB}\pd_A\Phi\pd_B\Phi-\tV(\Phi)\right),
\label{bulk_action}\\
S_{\text{brane}}&=&\int d^4x\sqrt{-g}\left(\tM^3\tF(\Phi)\Delta
K+L_{\text{brane}}(g_{\mu\nu},\Phi,\psi)\right).
\label{brane_action}
\end{eqnarray}
The bulk scalar field $\Phi$ is supposed to be a modulus field. This
is the field relevant to the hereby proposed modification of the DGP model.
$\tV(\Phi)$ may have a constant term $\tLambda$, being the
cosmological constant in the bulk. $\tF, \tX$ and $\tV$ do not
depend on $\tg$. $\Delta K=K^+-K^-$ is, as in (\ref{pure_DGP}), the jump of the trace of the
extrinsic curvature across the brane. Standard Model fields $\psi$
confined on the brane may also be present. They will not be needed in our
discussion. We choose to study the equations of motion (EOM) in the Gauss-Normal (GN) coordinate
system, with the brane
located at $y=0$. In GN coordinates it is convenient to identify
$X^\mu=\xi^\mu$ so that $x^\mu$ and $\xi^\mu$ are
indistinguishable (static gauge). Further, in this coordinate system $\tg=\diag(\tg_{\mu\nu},1)$
and the induced metric on the brane is
$g_{\mu\nu}=\tg_{\mu\nu}|_{y=0}$. The non-vanishing Christoffel symbols are
$\tGamma_{\mu\nu}^y=-\frac12\frac{\pd \tg_{\mu\nu}}{\pd
y},~\tGamma_{\mu y}^{\nu}=\frac12\tg^{\nu\rho}\frac{\pd
\tg_{\rho\mu}}{\pd y}$ and $\tGamma_{\mu\nu}^\rho$, where
$\tGamma_{\mu\nu}^\rho|_{y=0}={\Gamma}_{\mu\nu}^\rho$, the being the Christoffel
symbols obtained from the brane metric $g_{\mu\nu}$.
With $x^{\mu}$ and $\xi^\mu$ identified, we have $\pd_\mu=\pd/\pd x^\mu=\pd/\pd \xi^\mu$, while
$\pd_A\equiv \pd/\pd x^A$.

As a side remark, notice that if $\tF(\Phi)$ is not a constant, it can be
brought to any other non-constant form by an appropriate redefinition of the
field $\Phi$. The two actions are equivalent at the classical level.
A similar observation applies to the function $\tX(\Phi)$. If it is non-zero,
it may be transformed to any non-vanishing constant. The two redefinitions cannot, in general, be
applied together.


\subsection{Bulk}

Varying the bulk action (\ref{bulk_action}) with respect to the metric one obtains
in the bulk
\begin{equation}
\begin{split}
&-\tM^3\tF\tG_{AB}+\tM^3(\tD_A\tD_B\tF-\tg_{AB}\tBox\tF)+
\tX\pd_A\Phi\pd_B\Phi-\frac{\tX}2\tg_{AB}\tg^{CD}\pd_C\Phi\pd_D\Phi-{\tg_{AB}}\tV=0.\end{split}
\label{EOM_bulk_g}
\end{equation}
Here $\tG_{AB}=\tR_{AB}-\frac12\tR\tg_{AB}$ is the bulk Einstein
tensor, $\tD$ is the covariant derivative for the bulk metric
$\tg_{AB}$ and $\tBox=\tg^{AB}\tD_A\tD_B$. Similarly, the EOM of the scalar
field $\Phi$ reads
\begin{equation}
\tX\tBox\Phi+\frac{\tX^{(1)}}2\tg^{AB}\pd_A\Phi\pd_B\Phi+\frac{\tM^3\tF^{(1)}\tR}{2}-\tV^{(1)}=0.
\label{EOM_bulk_phi}
\end{equation}
Here the superscript ${}^{(n)}$ denotes the $n$-th derivative with respect to the field $\Phi$.

For simplicity we shall restrict ourselves to diagonal bulk metrics.
Furthermore, following the cosmological principle, we shall take
space on the brane to be homogeneous and isotropic. In the special
case of zero spatial curvature on the brane the most general ansatz
for the metric in GN coordinates is then
\begin{equation}
\tg_{AB}=\left(\begin{tabular}{ccc}$-N^2(t,y)$&0&0\\0&$A^2(t,y)\delta_{ab}$&0\\0&0&1\end{tabular}\right).
\label{metric_general}
\end{equation}
The non-zero Christoffel symbols for the $(\mu,\nu)$ part are
$\tGamma_{tt}^t=\dot N/N$, $\tGamma_{ab}^{t}=A\dot A \delta_{ab} /
N^2$, $\tGamma_{at}^b=\dot A\delta_a^b / A $. In what follows, we
will use dot for time derivative and prime for $y$ derivative.
Direct calculation of the Einstein tensor gives the following
non-zero components
\begin{eqnarray}
\tG_{tt}&=&(-N^2)3\left[-\frac{\dot
A^2}{A^2N^2}+\frac{{A'}^2}{A^2}+\frac{A''}{A}\right],\label{Gtt1}\\
\tG_{ab}&=&(A^2\delta_{ab})\left[-\frac{1}{N^2}\left(\frac{\dot
A^2}{A^2}+2\frac{\ddot A}A-2\frac{\dot A\dot
N}{AN}\right)+\frac{{A'}^2}{A^2}+2\frac{A''}A+2\frac{A'N'}{AN}+\frac{N''}N\right],\label{Gxx1}\\
\tG_{yy}&=&3\left[-\frac1{N^2}\left(\frac{\dot A^2}{A^2}+\frac{\ddot
A}A-\frac{\dot A\dot
N}{AN}\right)+\frac{{A'}^2}{A^2}+\frac{A'N'}{AN}\right],\label{Gyy1}\\
\tG_{ty}&=&3\left[\frac{\dot AN'}{AN}-\frac{\dot
A'}A\right].\label{Gty1}
\end{eqnarray}


\subsection{Brane}

Life on the brane is described by the equations of motion on the
brane. They are obtained from the brane lagrangian, supplemented by
the GH term. We use the following conventions: the unit vector $n^A$
normal to the brane is taken to point from the region $y<0$ into the
region $y>0$ in GN coordinates. The same on both sides of the brane.
In terms of this vector the induced metric is given by the tangent
to the brane components of the projection operator
$g_{AB}=\tg_{AB}-n_An_B$. The extrinsic curvature (the second
fundamental form of the surface) is defined as
$K_{AB}=-g_A^Cg_B^D\tD_Cn_D$ and its trace is $K=g^{AB}K_{AB}$. Its
components in the GN frame are
$K_{\mu\nu}=-\frac12\pd_y\tg_{\mu\nu}|_{y=\const} $. For our
purposes, we will have to evaluate it at $y=0+$ and $y=0-$, since we
put our brane at $y=0$ and allow for discontinuities of the $y$
derivatives of the metric components.

The equations of motion for the metric on the brane become
\begin{equation}
\begin{split}
&-\tM^3\left[\tF(K_{\mu\nu}-g_{\mu\nu}K)+g_{\mu\nu}\tF'\right]-
M^2FG_{\mu\nu}+M^2(D_\mu D_\nu F-g_{\mu\nu}\Box F)+\\
&+X\pd_\mu\Phi\pd_\nu\Phi-\frac{X}2g_{\mu\nu}g^{\alpha\beta}\pd_\alpha\Phi\pd_\beta\Phi-{g_{\mu\nu}}V=0.
\end{split}
\label{EOM_brane_g}
\end{equation}
Here $\Box=g^{\mu\nu}D_{\mu}D_{\nu}$ and $D_\mu$ is a covariant
derivative built upon the induced metric. From now on, we shall be
using square brackets to denote the discontinuity across the brane
of the quantity inside, i.e. $\left[W\right]\equiv W(y=0+)-W(y=0-)$,
for any quantity $W$. Of course, if one assumes $Z_2$ symmetry
across the brane, then $W(y=0+)=-W(y=0-)$ and
$\left[W\right]=2W(y=0+)$.

Similarly, the field $\Phi$ obeys the
following equation on the brane
\begin{equation}
\left[\tX'+\tM^3\tF^{(1)}K\right]+
X\Box\Phi+\frac{X^{(1)}}2g^{\mu\nu}\pd_\mu\Phi\pd_\nu\Phi+\frac{M^2F^{(1)}R}{2}-V^{(1)}=0.
\label{EOM_brane_phi}
\end{equation}


\section{Static scalar $\Phi=\Phi(y)$}

This is the simplest possibility, since to implement the
modification proposed in \cite{hep-th/0612213}, we need eventually a
non-trivial $y$-dependent function $\tF$. With $\Phi$ a function of
$y$ only, the system simplifies considerably. Namely,
$\Phi$-dependent quantities on the brane become constants and all
derivatives of $\Phi$ along the brane are zero. In the bulk only
$y$-derivatives survive and one is led to the following set of
equations.

In the bulk
\begin{eqnarray}
-\tM^3\tF\tG_{\mu\nu}+\tM^3\left(\frac12{\tg'}_{\mu\nu}\tF'-\frac12\tg_{\mu\nu}\bar
g \tF'-\tg_{\mu\nu}\tF''\right)
-\frac{\tX}2\tg_{\mu\nu}{\Phi'}^2-{\tg_{\mu\nu}}\tV&=&0,
\label{EOM_bulk_g_y_munu2}\\
-\tM^3 \tF G_{yy}-\frac{\tM^3}2\bar g \tF'
+\frac{\tX}2{\Phi'}^2-\tV&=&0,
\label{EOM_bulk_g_y_yy2}\\
\tX\left(\Phi''+\frac12\bar g
\Phi'\right)+\frac{\tX^{(1)}}2{\Phi'}^2+\frac{\tM^3\tF^{(1)}\tR}{2}-\tV^{(1)}&=&0,
\label{EOM_bulk_phi_y2}\\
\frac{\dot AN'}{AN}-\frac{\dot A'}A&=&0.\label{Ryyy1_1}
\end{eqnarray}
Where $\bar g\equiv \tg^{\alpha\beta}\tg'_{\alpha\beta}$ and $\tBox=\pd_y^2+\frac12\bar g\pd_y$.
Contracting the Einstein equations (\ref{EOM_bulk_g_y_munu2}) and (\ref{EOM_bulk_g_y_yy2})
with $\tg^{AB}$ one obtains
\begin{equation}
\frac{3\tM^3\tF\tR}2-2\tM^3\left({\bar g}\tF'+2\tF''\right)
-\frac{3\tX}2{\Phi'}^2-5\tV=0. \label{bulk_trick}
\end{equation}
Under our assumption that $\tF'\neq0$ we can solve
(\ref{EOM_bulk_phi_y2}) and (\ref{bulk_trick}) for $\tR$ and $\bar g$, provided
\begin{equation}
\tX\neq -{\frac43}\tM^3\frac{(\tF^{(1)})^2}{\tF}. \label{condition}
\end{equation}
This leads to time-independent $\tR$ and $\bar g$. In what follows
we assume that condition (\ref{condition}) is valid.

On the brane the field equations are
\begin{eqnarray}
-\tM^3g_{\mu\nu}[\tF']-\tM^3\tF \left[K_{\mu\nu}-g_{\mu\nu}K\right]-
M^2FG_{\mu\nu}-{g_{\mu\nu}}V&=&0,
\label{EOM_brane_g_y}\\
{}[\tX']+\tM^3\tF^{(1)}
\left[K\right]+\frac{M^2F^{(1)}R}{2}-V^{(1)}&=&0.
\label{EOM_brane_phi_y}
\end{eqnarray}
All tilded quantities are evaluated on the brane and consequently
are constants. Contracting equation (\ref{EOM_brane_g_y}) with
$g^{\mu\nu}$ one obtains
\begin{equation}
-4\tM^3[\tF']+3\tM^3\tF \left[K\right]+ M^2FR-4V=0.
\label{brane_trick}
\end{equation}
This, together with (\ref{EOM_brane_phi_y}) form a system of two linear
algebraic equations for $[K]$ and $R$, which has a
solution provided $3\tF F^{(1)}-2\tF^{(1)}F\neq0$. In this case both
$[K]$ and $R$ are constants. If, on the other hand, $3\tF
F^{(1)}-2\tF^{(1)}F=0$, then for consistency we have to require $3\tF
\left(V^{(1)}-[\tX']\right)=4\tF^{(1)}\left(V+\tM^3[\tF']\right)$
and $[K]$ and $R$ would not, a priori, have to be
time-independent. However, given that in our notation
$K(y_0)=-\frac12g^{\mu\nu}g'_{\mu\nu}=-\frac12\tg|_{y=y_0}$, which is time-independent,
we conclude that $R$ is also necessarily time-independent, a consequence
of our assumption that $\Phi=\Phi(y)$.

It is rather straightforward to take into account the remaining equations. One is
led to the following conclusions:
\begin{itemize}
\item $\tR$, $\bar g$, $R$ and $[K]$ do not depend on time
\item $A=\alpha(y)a(t)$ and $N=\nu(y)n(t)$, i.e. the functions $A$ and $N$ are factorized.
\item There are two possibilities for these functions:
either $N=b(t)\dot A$, or $\dot A=0$ and $N$ undetermined.
\end{itemize}
The second possibility gives
Minkowski brane, which is not particularly interesting for our purposes.
The solution in this case can be found analytically, reduced to quadratures.
The first possibility, to which we concentrate next,
looks more interesting for cosmology, because it leads to a time-dependent metric
on the brane.


\section{Construction of the solution for static scalar and $N=b(t)\dot A$}


\subsection{Bulk part}

In this case $n(t)=b(t)\dot a(t)$ and $\alpha(y)=\nu(y)$.
The non-zero components of the Einstein tensor become
\begin{equation*}
\begin{split}
\tG_{tt}&=(-b^2\dot A^2)3P,~
\tG_{ab}=(A^2\delta_{ab})\left(P+2Q\right),~
\tG_{yy}=3\left(P+Q-2\frac{\nu''}{\nu}\right),\\
~\text{where}~P&=-\frac{1}{A^2b^2}+\frac{{\nu'}^2}{\nu^2}+\frac{\nu''}{\nu},
~Q=\frac{\dot b}{A\dot
Ab^3}+\frac{{\nu'}^2}{\nu^2}+\frac{\nu''}{\nu}.
\end{split}
\end{equation*}
Let us, now, take a closer look at the $\{tt\}$ component of
equation (\ref{EOM_bulk_g_y_munu2}). All terms in this equation,
with the exception of the first one, are of the form $\dot
a\times{\rm {(a\; function \;of\; y)}}$. This is a direct
consequence of the  factorization property of $\tg_{tt}$ and the
$t-$independence of $\Phi$. Only the first term may a priori depend
on time, due to the presence of the $-1/({A^2b^2})$ term in $P$.
Thus, the $\{tt\}$ component of equation (\ref{EOM_bulk_g_y_munu2})
for non-vanishing $\dot a$ takes the form $a(t)b(t)={\rm
{(function\; of \;y)}}$, and leads to $b=1/(Ha)$ where $H$ is a
constant. The Einstein tensor then becomes
\begin{equation}
\tG_{\mu\nu}=\frac3{\nu^2}(-H^2+{\nu'}^2+\nu\nu'')\tg_{\mu\nu},\quad
\tG_{yy}=\frac6{\nu^2}(-H^2+{\nu'}^2).
\end{equation}

The Einstein equations read
\begin{eqnarray}
\frac3{\nu^2}(-H^2+{\nu'}^2-\nu\nu'')&=&
+\frac{1}{\tM^3\tF}\left(\tX{\Phi'}^2+\tM^3\tF''-\tM^3\frac{\nu'}\nu\tF'\right),
\label{Emunuy1}\\
\frac6{\nu^2}(-H^2+{\nu'}^2)&=&-\frac{1}{\tM^3\tF}\left(-\frac12\tX{\Phi'}^2+\tV
+4\tM^3\frac{\nu'}\nu\tF'\right). \label{Eyyy1}
\end{eqnarray}
Notice that $a(t)$ remains arbitrary and will not be fixed by the
equations of motion. However, any form of $a(t)$ leads to (anti) de
Sitter brane metric, because it can be  absorbed into the definition
of time by $\dot a/(HA)dt=d\tau$, in terms of this new time $\tau$
one obtains $a(\tau)=a_0e^{H\tau}$. So, the bulk metric reads
\begin{equation}
\tg_{AB}=\left(\begin{tabular}{cc}$\nu^2(y)\left(\begin{tabular}{cc}$-1$&$0$\\$0$&$e^{2H\tau}$\end{tabular}\right)$&$0$\\$0$&1\end{tabular}\right)
\label{sol_tg_y}
\end{equation}
where, without loss of generality, $a_0$ was chosen equal to $1$.

Given the functions $\tF(y)$ and $\nu(y)$ one can use (\ref{Emunuy1}) to obtain
$\tX{\Phi'}^2$ as a function of $y$
\begin{equation}
\tX{\Phi'}^2=\frac{3h\tM^3\tF}{\nu^2}-\tM^3\left(\tF''-\frac{\nu'}\nu\tF'\right)
\label{xphiy2ysol}
\end{equation}
where $h=-H^2+{\nu'}^2-\nu\nu''$. Finally, plug this into
equation (\ref{Eyyy1}) and express the potential $\tV$
\begin{equation}
\tV=-\frac{(9h+12\nu\nu'')\tM^3\tF}{2\nu^2}-\frac{\tM^3}2\left(\tF''+\frac{7\nu'}\nu\tF'\right).
\label{vysol}
\end{equation}
One can check by direct substitution that equation
(\ref{EOM_bulk_phi}) is satisfied.


\subsection{Life on the brane}

The geometry and dynamics on the brane, using the time coordinate $\tau$, are given by
\begin{equation*}
\begin{split}
g_{\mu\nu}&=\left(\begin{tabular}{cc}$-1$&$0$\\$0$&$e^{2H\tau}\delta_{ab}$\end{tabular}\right),~
R={12H^2},~G_{\mu\nu}=-{3H^2}g_{\mu\nu},\\
[K_{\mu\nu}]&=-[\nu']g_{\mu\nu},~ [K]=-4[\nu'].
\end{split}
\end{equation*}
As usual, $[\nu']\equiv \nu'_+ - \nu'_-$ is the discontinuity of the
function $\nu'(y)$ across the brane. $Z_2$ symmetry would force them
to satisfy $\nu'_+=-\nu'_-$. Also we choose $\nu(0)=1$.

As discussed above, we have to distinguish two cases.  (a) If $3\tF
F^{(1)}-2\tF^{(1)}F\neq0$ we may unambiguously express $[\nu']$ and
$H$ through $\tF$, $F$ and $V$ as follows
\begin{equation}
\begin{split}
[\nu']&=\frac{F\left(V^{(1)}-[\tX']\right)-2F^{(1)}\left(V+\tM^3[\tF']\right)}{2\tM^3(3\tF F^{(1)}-2\tF^{(1)}F)},\\
{H}&=\pm\sqrt{\frac{3\tF
\left(V^{(1)}-[\tX']\right)-4\tF^{(1)}\left(V+\tM^3[\tF']\right)}{6M^2(3\tF
F^{(1)}-2\tF^{(1)}F)}}.
\end{split}
\label{sol_y}
\end{equation}
(b) If, on the other hand, $3\tF F^{(1)}-2\tF^{(1)}F=0$, we have to
require $3\tF \left(V^{(1)}-[\tX']\right)=
4\tF^{(1)}\left(V+\tM^3[\tF']\right)$. In this case, one of the
quantities $[\nu']$ and $H$ remains undetermined. However, they are
related by
\begin{equation}
\begin{split}
H&=\pm\sqrt{\frac{\tM^3\tF}{M^2F}[\nu']+\frac{V+\tM^3[\tF']}{3M^2F}}.
\end{split}
\label{sol_y0}
\end{equation}
As a special case, it is easy to check that taking $\tF=F=1$, $\tX=X=\tV=0$, $V=\sigma$
and $\nu'_+=-\nu'_-=\varepsilon H$ one reproduces the known
relation (\ref{Hwithsigma}) between brane-tension and $H$ of the DGP setup.

\section{Application to the modified DGP model and discussion}

Let us recapitulate the steps one has to take to construct a
modified model of the type we are proposing here, based on $\Phi$
which depends only on $y$. For our purposes, it is convenient to
start with a given $\tF$, since the modified DGP model
\cite{hep-th/0612213} gives some idea about the desirable form of
$\tF(y)$. Namely, $\tF=1-m\bar\delta(y)$ such that
$\tF(y=0)=\const<0$ and $\tF(y\gg\alpha)=1$, where $\alpha$ is the
width of localization of $\tF$ near the brane. The metric has the
general form (\ref{sol_tg_y}). Given, in addition, the function
$\nu(y)$, one uses equations (\ref{xphiy2ysol}) and (\ref{vysol}) to
obtain $\tX$ and $\tV$ as functions of $y$. Finally, on the brane,
one has to satisfy (\ref{EOM_brane_g_y}) and
(\ref{EOM_brane_phi_y}). It remains to determine the $\Phi$
dependence of the bulk quantities. For that, one needs to specify
(or know independently) either $\Phi(y)$ or the dependence on $\Phi$
of one of $\tF$, $\tX$ or $\tV$.

Consider the special case of Minkowski bulk. This fixes
$\nu(y)=1+\nu_1 y$, with $\nu_1=\varepsilon H$ and we assume $Z_2$
symmetry. Take
$\tF=1-\frac{m}{\alpha\cosh\left({y^2}/{\alpha^2}\right)}$ and
$\Phi(y)=y$. One is led to the following model
\begin{equation}
\begin{split}
\tF&=1-\frac{m}{\alpha\cosh\left({\Phi^2}/{\alpha^2}\right)},\\
\tX&=-\frac{m\tM^3}{\alpha^5\cosh^3\left({\Phi^2}/{\alpha^2}\right)}
\left(2\Phi^2\left(3-\cosh\left(2{\Phi^2}/{\alpha^2}\right)\right)
+\alpha^2\frac{\sinh\left(2{\Phi^2}/{\alpha^2}\right)}{1+\nu_1\Phi}\right),\\
\tV&=-\frac{m\tM^3}{\alpha^5\cosh^3\left({\Phi^2}/{\alpha^2}\right)}
\left(\Phi^2\left(3-\cosh\left(2{\Phi^2}/{\alpha^2}\right)\right)
+\alpha^2\frac{1+8\nu_1\Phi}{2(1+\nu_1\Phi)}\sinh\left(2{\Phi^2}/{\alpha^2}\right)\right),\\
F&=1,\qquad V=\sigma.
\end{split}
\label{explicit}
\end{equation}
It is easy to check that bulk equations are satisfied. Equation
(\ref{EOM_brane_phi_y}) on the brane is satisfied trivially, while
equation (\ref{EOM_brane_g_y}) reduces to
\begin{eqnarray}
-6\tM^3\left(1-\frac{m}{\alpha}\right) \varepsilon H+
3M^2H^2-\sigma&=&0 \label{EOM_brane_g_y0}
\end{eqnarray}
which determines $H$ as
\begin{equation}
H=\frac{\tM^3}{M^2}\left(1-\frac{m}{\alpha}\right)\left(\varepsilon\pm\sqrt{1+\frac{\sigma
M^2}{3\tM^6\left(1-\frac{m}{\alpha}\right)^2}}\right) \label{Hour}
\end{equation}
which coincides with one given in (\ref{Hgabadadge}) up to the
rescaling of $\alpha$ by $\pi$ and thus explicitly realizes the
trick outlined in \cite{hep-th/0612213}.

As another special model, let us take $\tX=0$, still with
Minkowski bulk. Then expression (\ref{xphiy2ysol}) becomes an
equation for the function $\tF$ with the general solution
\begin{equation}
\tF(y)=C_1(1+\nu_1y)^2+C_2. \label{tFtX0}
\end{equation}
Here we see that $\tF$ is not localized near the brane and becomes
important in the whole range of $y$ up to the Rindler horizon
$y=-1/\nu_1$. However, the potentially dangerous domain of $y$ with
$\tF<0$ can be made arbitrarily narrow, with $|\tF(y=0)|$ also
infinitesimal. Indeed, adjust $\tF(y=0)=C_2+C_1=-\alpha<0$ to be
extremely small. Then, the width of the negative domain becomes
$-\alpha/(C_2\nu_1)$. In order to have realistic $H$ we have to
adjust ($\alpha\tM^3)\sim(100\text{MeV})^3$ \cite{gg}.

Let us briefly summarize the results of the present analysis. A
covariant implementation of the modified DGP model
\cite{hep-th/0612213} was constructed, by means of a bulk scalar
field, with appropriate coupling to gravity. Flat metric in the bulk
is a possible solution with non-constant scalar field. The
parameters of the theory can be adjusted in such a way that one has
not growing metric in the bulk (the usual normal-branch behavior)
with positive Hubble constant on the brane, leading to accelerating
expansion on the brane, even without any extra matter and even with
negative brane tension. We explicitly demonstrated two models
satisfying our requirements (a)-(d) of the Introduction. (i) In the
first one the $\delta$-function-like profile of the function $\tF$
is localized near the brane and in the limit $\alpha\to0$ the width
of localization goes to zero. Thus, in the $\alpha\to0$ limit the
bulk is modified in a narrow domain near the brane only. However,
the presence of the kinetic term for the scalar field in the bulk is
necessary. (ii) In the second model, on the other hand, the scalar
field does not have a kinetic term, but the profile of $\tF$ does
not go rapidly to unity away from the brane. However, one may adjust
integration constants and make the $y$-domain with negative $\tF$
infinitesimal, if necessary to be consistent with observations.

A detailed stability analysis of the hereby proposed class of models
is an open question. Also, time-dependent bulk scalar field
solutions, which were not studied here, may give rise to interesting
phenomena on the brane. Finally, asymmetric brane-world setups in
the spirit of the recent paper \cite{recent} may provide further
possibilities for model building.


\section*{Acknowledgements}
Authors are grateful to D.Gal'tsov, R.Gregory, G.Kofinas, A.Petkou
and M.Smolyakov for useful comments and discussions. The work is
supported in part by EU grants MRTN-CT-2004-512194 and Marie Curie Fellowship MIF1-CT-2005-021982.
A.K. is also supported
in part by RFBR grant 05-01-00758, INTAS grant 03-51-6346 and Russian President's grant
NSh-2052.2003.1.


\end{document}